\documentclass[a4paper,11pt]{article}
\usepackage{jheppub,graphicx,amssymb,cancel}

\def\be{\begin{equation}}
\def\ee{\end{equation}}
\def\bea{\begin{eqnarray}}
\def\eea{\end{eqnarray}}
\def\nnb{\nonumber}

\newcommand{\scs}{\scriptscriptstyle}
\newcommand{\f}{\frac}
\newcommand{\fm}[2]{{\textstyle \frac{#1}{#2}}}
\newcommand{\al}{\alpha_{\mathrm s}}

\newcommand{\ep}{\epsilon}
\newcommand{\G}{\hat{G}}

\title{\boldmath Towards $\bar B \to X_s \gamma$ at the NNLO in QCD without interpolation in $m_c$}

\author[a,1]{Miko{\l}aj Misiak,\note{Corresponding author.}}
\author[b,c]{Abdur Rehman,}
\author[b]{and Matthias Steinhauser}

\affiliation[a]{Institute of Theoretical Physics, Faculty of Physics,University of Warsaw,\\ 
                Pasteura 5, 02-093 Warsaw, Poland.}
\affiliation[b]{Institut f\"ur Theoretische Teilchenphysik, Karlsruhe Institute of Technology (KIT),\\
                76128 Karlsruhe, Germany.}
\affiliation[c]{National Centre for Physics, Quaid-i-Azam University Campus,\\
                Islamabad 45320, Pakistan.}

\emailAdd{Mikolaj.Misiak@fuw.edu.pl}
\emailAdd{Abdur.Rehman@ncp.edu.pk}
\emailAdd{Matthias.Steinhauser@kit.edu}

\abstract{
Strengthening constraints on new physics from the $\bar B \to X_s \gamma$
branching ratio requires improving accuracy in the measurements and
the Standard Model predictions. To match the expected Belle-II
accuracy, Next-to-Next-to-Leading Order (NNLO) QCD corrections must be
calculated without the so-far employed interpolation in the
charm-quark mass $m_c$. In the process of evaluating such corrections
at the physical value of $m_c$, we have finalized the part coming from
diagrams with closed fermion loops on the gluon lines that contribute
to the interference of the current-current and photonic dipole
operators. We confirm several published results for corrections of
this type, and supplement them with a previously uncalculated
piece. Taking into account the recently improved estimates of 
non-perturbative contributions, we find 
${\mathcal B}_{s \gamma} = (3.40 \pm 0.17) \times 10^{-4}$~
and~
$R_\gamma \equiv {\mathcal B}_{(s+d) \gamma}/{\mathcal B}_{c\ell\bar\nu} = (3.35 \pm 0.16) \times 10^{-3}$~
for~ $E_\gamma > 1.6\,$GeV in the decaying meson rest frame.}

\arxivnumber{2002.01548}

\begin{document}
\maketitle

\section{Introduction \label{sec:intro}}

Flavour Changing Neutral Current (FCNC) processes receive the leading
Standard Model (SM) contributions from one-loop diagrams only, often with
additional suppression factors originating from the
Glashow-Iliopoulos-Maiani (GIM) mechanism~\cite{Glashow:1970gm}. It
makes them sensitive to possible existence of new weakly-interacting
particles with masses ranging up to ${\mathcal O}(100\,{\rm
TeV})$. Significant deviations from the SM predictions are observed in
the GIM-unsuppressed FCNC processes mediated by the $b \to s \mu^+
\mu^-$ transition (see, e.g., the recent summary in
Ref.~\cite{Aebischer:2019mlg}). On the other hand, no deviations are
seen in the closely related $b \to s \gamma$ transition, despite
higher accuracy of both the measurements and the SM predictions in its
case.

The physical observable giving the strongest constraints on the $b \to
s \gamma$ amplitude is the inclusive ${\mathcal B}_{s \gamma}$
branching ratio, i.e.\ the CP- and isospin- averaged branching ratio of
$\bar B \to X_s \gamma$ and $B \to X_{\bar s} \gamma$ decays, with
$\bar B$ and $B$ denoting ($\bar B^0$ or $B^-$) and ($B^0$ or $B^+$),
respectively. The states $X_s$ and $X_{\bar s}$ are assumed to contain
no charmed hadrons.  ${\mathcal B}_{s \gamma}$ is being
measured~\cite{Chen:2001fja,Aubert:2007my,Lees:2012wg,Lees:2012ym,Saito:2014das,Belle:2016ufb}
with $E_\gamma > E_0$ for $E_0 \in [1.7,2.0]$\,GeV, and then
extrapolated to the conventionally chosen value of $E_0 = 1.6\,$GeV
to compare with the theoretical predictions (that would be less
accurate at higher $E_0$). The current experimental world average for
${\mathcal B}_{s \gamma}$ at $E_0 = 1.6\,$GeV reads $(3.32 \pm 0.15)
\times 10^{-4}$~\cite{Amhis:2019ckw}, which corresponds to an
uncertainty of around $\pm 4.5\%$. With the full Belle-II dataset, the
world average uncertainty at the level of $\pm 2.6\%$ is
expected~\cite{Kou:2018nap,Ishikawa:2019Lyon}.  Achieving a similar
accuracy in the SM predictions is essential for improving the power of
${\mathcal B}_{s \gamma}$ as a constraint on Beyond-SM (BSM) theories. It is
the goal of the calculations we describe in what follows.

The SM prediction for ${\mathcal B}_{s \gamma}$ (see
Refs.~\cite{Misiak:2015xwa,Czakon:2015exa}), is based on the formula
\be \label{brB}
{\mathcal B}(\bar B \to X_s \gamma)_{E_{\gamma} > E_0}
= {\mathcal B}(\bar B \to X_c \ell \bar \nu)
\left| \f{ V^*_{ts} V_{tb}}{V_{cb}} \right|^2 
\f{6 \alpha_{\mathrm em}}{\pi\;C} 
\left[ P(E_0) + N(E_0) \right],
\ee
where $\alpha_{\mathrm em} = \alpha_{\mathrm em}^{\rm on\, shell}$,
while the so-called semileptonic phase-space factor $C$ is given by
\be \label{phase}
C = \left| \f{V_{ub}}{V_{cb}} \right|^2 
\f{\Gamma[\bar{B} \to X_c e \bar{\nu}]}{\Gamma[\bar{B} \to X_u e \bar{\nu}]}.
\ee
Its numerical value is determined~\cite{Alberti:2014yda} using the Heavy Quark Effective Theory (HQET) methods
from measurements of the $\bar B \to X_c \ell \bar \nu$ decay spectra.
The quantity $P(E_0)$ is defined through the following ratio of
perturbative inclusive decay rates of the $b$ quark:
\be \label{pert.ratio}
\f{\Gamma[ b \to X^p_s \gamma]_{E_{\gamma} > E_0}}{
|V_{cb}/V_{ub}|^2 \; \Gamma[ b \to X^p_u e \bar{\nu}]} ~=~ 
\left| \f{ V^*_{ts} V_{tb}}{V_{cb}} \right|^2 
\f{6 \alpha_{\mathrm em}}{\pi} \; P(E_0),
\ee
with $X^p_s$ and $X^p_u$ denoting all the possible charmless partonic
final states in the respective decays ($X^p_s = s, sg, sq\bar q, \ldots$).
The non-perturbative contribution from $N(E_0)$ in Eq.~(\ref{brB}) is estimated\footnote{
See Sec.~\ref{sec:num} for details on the current uncertainty budget.}
at the level of around $4\%$ of ${\mathcal B}_{s \gamma}$.
%
%
To achieve ${\mathcal O}(3\%)$ precision in $P(E_0)$, evaluation of
the Next-to-Next-to Leading (NNLO) QCD corrections to this quantity is necessary.
\begin{figure}[t]
\begin{tabular}{lll}
\hspace{-2mm} \includegraphics[scale=0.54]{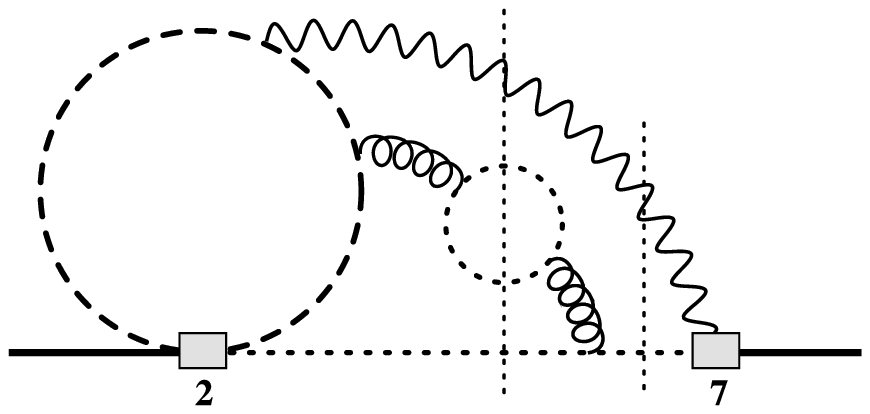} 
           &  \includegraphics[scale=0.54]{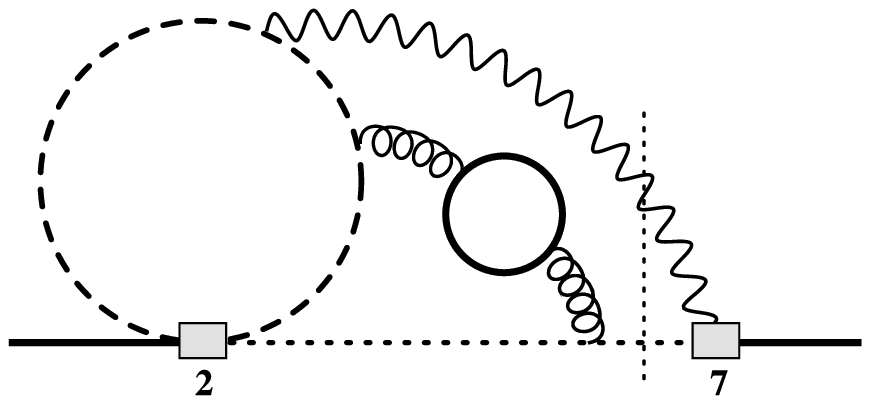} 
           &  \includegraphics[scale=0.54]{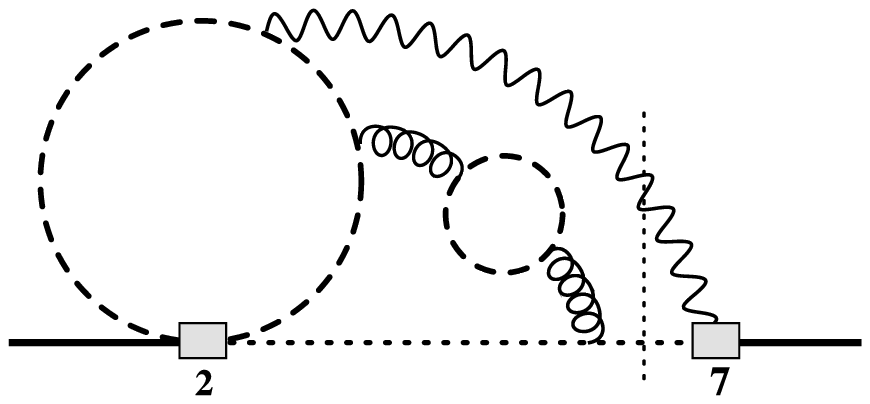}\\
\end{tabular}
\begin{tabular}{lll}
\hspace{-2mm} \includegraphics[scale=0.54]{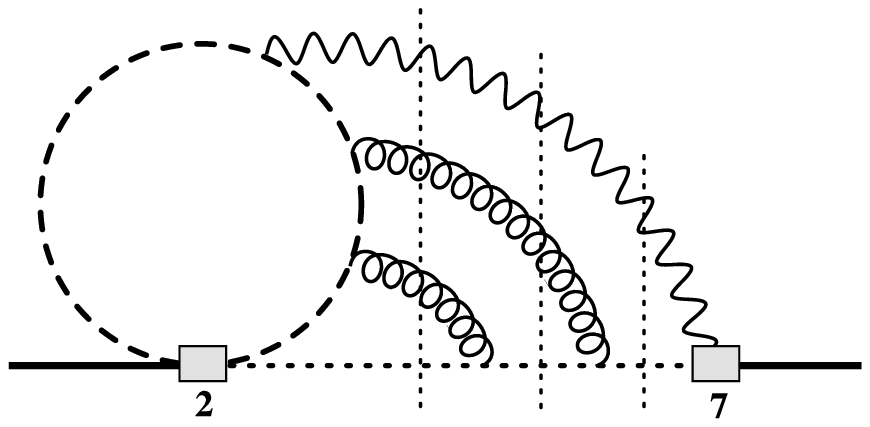}  
           &  \includegraphics[scale=0.54]{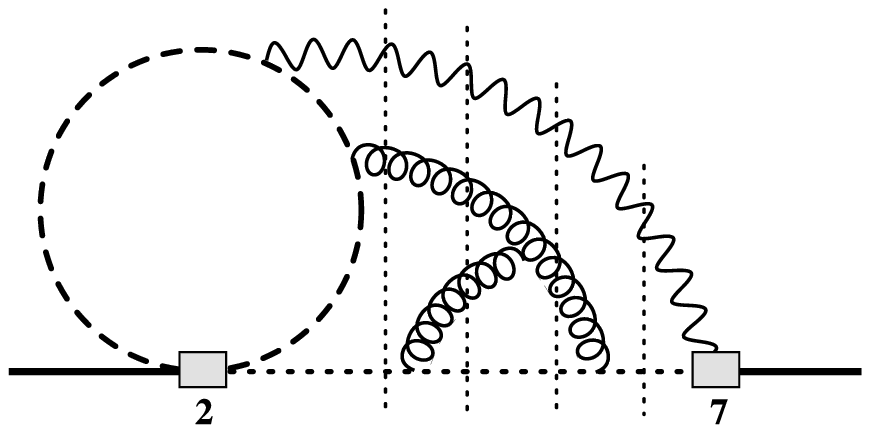}
           &  \includegraphics[scale=0.54]{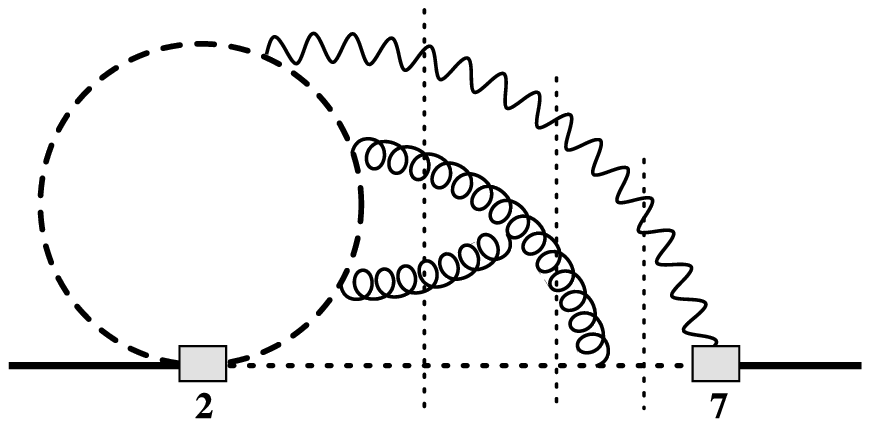}\end{tabular}
\caption{ \sf Sample Feynman diagrams contributing to $\G^{(2)}_{27}$ at ${\mathcal O}(\al^2)$. The vertical
dotted lines indicate possible unitarity cuts. The dotted, dashed and solid propagators
correspond to quarks with masses $0$, $m_c$ and $m_b$, respectively.} \label{fig:sample.diagrams}
\end{figure}

Perturbative calculations of $P(E_0)$ are most conveniently performed
in the framework of an effective theory obtained from the SM via
decoupling of the $W$ boson and all the heavier particles. The
relevant weak interactions are then given by the following Lagrangian
density\footnote{
For simplicity, we refrain here from displaying those terms in ${\mathcal L}_{\rm
weak}$ that matter for subleading electroweak or CKM-suppressed effects only.
Such effects have been included in the numerical analysis of Refs.~\cite{Misiak:2015xwa,Czakon:2015exa}.}
\be
{\mathcal L}_{\rm weak} = \f{4G_F}{\sqrt{2}} V_{ts}^\star V_{tb} \sum_{i=1}^8 C_i(\mu_b) Q_i.
\ee
Evaluation of the Wilson coefficients $C_i$ to the NNLO accuracy
$\left({\mathcal O}(\al^2)\right)$ at the renormalization scale $\mu_b \sim m_b$
required computing electroweak-scale matching up to three
loops~\cite{Misiak:2004ew}, and QCD anomalous dimensions up to four
loops~\cite{Czakon:2006ss}. Since $C_i$ in the SM have no imaginary
parts, one can write the perturbative decay rate as
\be \label{def.ghat}
\Gamma(b \rightarrow X^p_s \gamma) = 
\f{G_F^2\, m^5_{b,\,{\rm pole}}\,\alpha_{\mathrm em}}{32\pi^4} \left| V_{ts}^* V_{tb} \right|^2
\sum_{i,j=1}^8 C_i(\mu_b) C_j(\mu_b) \G_{ij},\hspace{15mm} (\G_{ij} = \G_{ji}),
\ee
where $\G_{ij}$ come from interferences of amplitudes with
insertions of the operators $Q_i$ and $Q_j$. The dominant NNLO effects
come from $\G_{17}$, $\G_{27}$ and $\G_{77}$ that originate from the
operators
\be \label{ops}
Q_1  = (\bar{s}_L \gamma_{\mu} T^a c_L) (\bar{c}_L     \gamma^{\mu} T^a b_L),\hspace{8mm}
Q_2  = (\bar{s}_L \gamma_{\mu}     c_L) (\bar{c}_L     \gamma^{\mu}     b_L),\hspace{8mm}
Q_7  =  \fm{em_b}{16\pi^2} (\bar{s}_L \sigma^{\mu \nu}     b_R) F_{\mu \nu}.
\ee
Whereas $\G_{77}$ has been known up to ${\mathcal O}(\al^2)$ since a long
time~\cite{Blokland:2005uk,Melnikov:2005bx,Asatrian:2006ph,Asatrian:2006sm,Asatrian:2006rq},
no complete NNLO calculation of $\G_{17}$ and $\G_{27}$ at the
physical value of the charm quark mass $m_c$ has been finalized to
date. Instead, calculations of these quantities at $m_c \gg m_b$~\cite{Misiak:2006ab,Misiak:2010sk}
and $m_c=0$~\cite{Czakon:2015exa} gave a basis for estimating their physical values
using interpolation~\cite{Czakon:2015exa}. The related uncertainty in ${\mathcal B}_{s
\gamma}$ (due to the $m_c$-interpolation only) has been estimated at the
level of $\pm 3\%$, which places it among the dominant contributions
to the overall theoretical uncertainty (see Sec.~\ref{sec:num}).

To calculate the interferences $\G_{ij}$ at the physical value of
$m_c$, it is convenient to express them in terms of propagator
diagrams with unitarity cuts. Examples of such four-loop diagrams
contributing to $\G_{27}$ at ${\mathcal O}(\al^2)$ are shown in
Fig.~\ref{fig:sample.diagrams}, with the light quarks ($u$, $d$, $s$)
treated as massless. Similar diagrams for $\G_{17}$ differ from the
$\G_{27}$ ones by simple colour factors only. For definiteness, we
shall focus on $\G_{27}$ in what follows.
\begin{figure}[t]
\begin{center} \includegraphics[width=13cm]{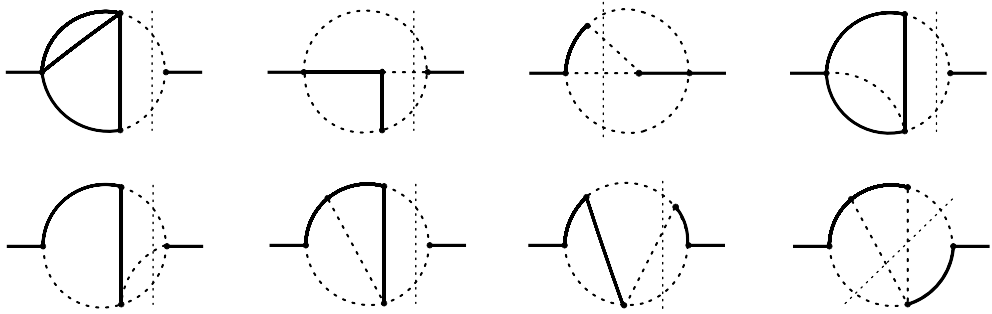} \end{center}
\caption{\sf Sample three-loop propagator-type integrals that
parameterize large-$z$ expansions of the MIs. Massless and massive
internal propagators are denoted by dotted and solid lines,
respectively. The thin dotted lines indicate the unitarity cuts.}
\label{fig:SSIs}
\end{figure}

By analogy to what has been done in the $\G_{77}$
case~\cite{Blokland:2005uk,Melnikov:2005bx,Asatrian:2006ph,Asatrian:2006sm,Asatrian:2006rq},
evaluation of ${\mathcal O}(\al^2)$ contributions to $\G_{27}$ is
performed in two steps.  First, no restriction on the photon energy $E_\gamma$ is
assumed. Next, one performs the calculation for $E_\gamma <
E_0$, which requires considering diagrams with three- and four-body
cuts only.  The desired result
$\G_{27}^{E_\gamma > E_0} = \G_{27}^{{\rm any}\, E_\gamma} - \G_{27}^{E_\gamma < E_0}$
is then obtained without necessity of determining the differential
photon spectrum close to the endpoint $E_\gamma^{\rm max} = \f12 m_b$.

In the present paper, we describe our calculation of $\G_{27}^{(2)}$ in
\be
\G_{27} ~=~ \f{\al}{4\pi}\, \G_{27}^{(1)} ~+~ \left(\f{\al}{4\pi}\right)^2 \G_{27}^{(2)} ~+~ {\mathcal O}(\al^3)
\ee
at the physical value of $m_c$, and with no restriction on $E_\gamma$.
Final results are presented for contributions originating from
diagrams with closed fermion loops on the gluon lines, like those in
the first row of Fig.~\ref{fig:sample.diagrams}.  They undergo
separate renormalization and are gauge invariant on their own, so they
serve as a useful test case for our calculation of the complete $\G_{27}^{(2)}$.
Most of such contributions have already been determined in the
past~\cite{Ligeti:1999ea,Bieri:2003ue,Boughezal:2007ny,Misiak:2010tk}
and implemented in the phenomenological analysis~\cite{Misiak:2015xwa,Czakon:2015exa}.
We confirm the published results, and supplement them with a
previously uncalculated piece. Some of the previous results have been
obtained by a single group only, which makes our verification relevant.

The article is organized as follows. In the next section, our
algorithm for evaluation of the complete $\G_{27}^{(2)}$ is
sketched, and the current status of the calculation is
summarized. Next, we focus on the closed fermionic loop contributions,
displaying our numerical results and comparing them with the
literature wherever possible. In Sec.~\ref{sec:num}, the SM prediction
for the branching ratio is updated, taking into account the recently
improved estimates of non-perturbative effects~\cite{Gunawardana:2019gep}.
We conclude in Sec.~\ref{sec:summary}. In the Appendix, large-$z$ expansions
of our final results are presented, and one of the counterterm
contributions is discussed.

\section{\boldmath The NNLO contribution to $\G_{27}$ \label{sec:pert}}

The quantity $\G_{27}^{(2)}$ is given by a few hundreds of
four-loop propagator diagrams with unitarity cuts, as those presented
in Fig.~\ref{fig:sample.diagrams}. We generate them using {\tt QGRAF}~\cite{Nogueira:1991ex}
and/or {\tt FeynArts}~\cite{Kublbeck:1990xc,Hahn:2000kx}. After
performing the Dirac algebra with the help of {\tt FORM}~\cite{Ruijl:2017dtg},
we express the full $\G_{27}^{(2)}$ in terms of several hundred
thousands scalar integrals grouped in ${\mathcal O}(500)$
families.\footnote{
Integrals in a family differ only by indices, i.e.\ the powers to which the propagators and/or irreducible numerators are being raised.}
Next, the Integration By Parts (IBP) identities~\cite{Tkachov:1981wb,Chetyrkin:1981qh,Laporta:2001dd}
for each family are generated and applied using {\tt KIRA}~\cite{Maierhoefer:2017hyi,Maierhofer:2018gpa},
as well as {\tt FIRE}~\cite{Smirnov:2014hma,Smirnov:2019qkx} and
{\tt LiteRed}~\cite{Lee:2012cn,Lee:2013mka}. In effect, $\G_{27}^{(2)}$ becomes a
linear combination of Master Integrals (MIs). The IBP reduction is the
most computer-power demanding part of the calculation, with ${\mathcal
O}(1\,{\rm TB})$ RAM nodes and weeks of CPU time needed for the most
complicated families.
\begin{figure}[t]
\begin{center}
\begin{tabular}{ccc}
\includegraphics[scale=0.52]{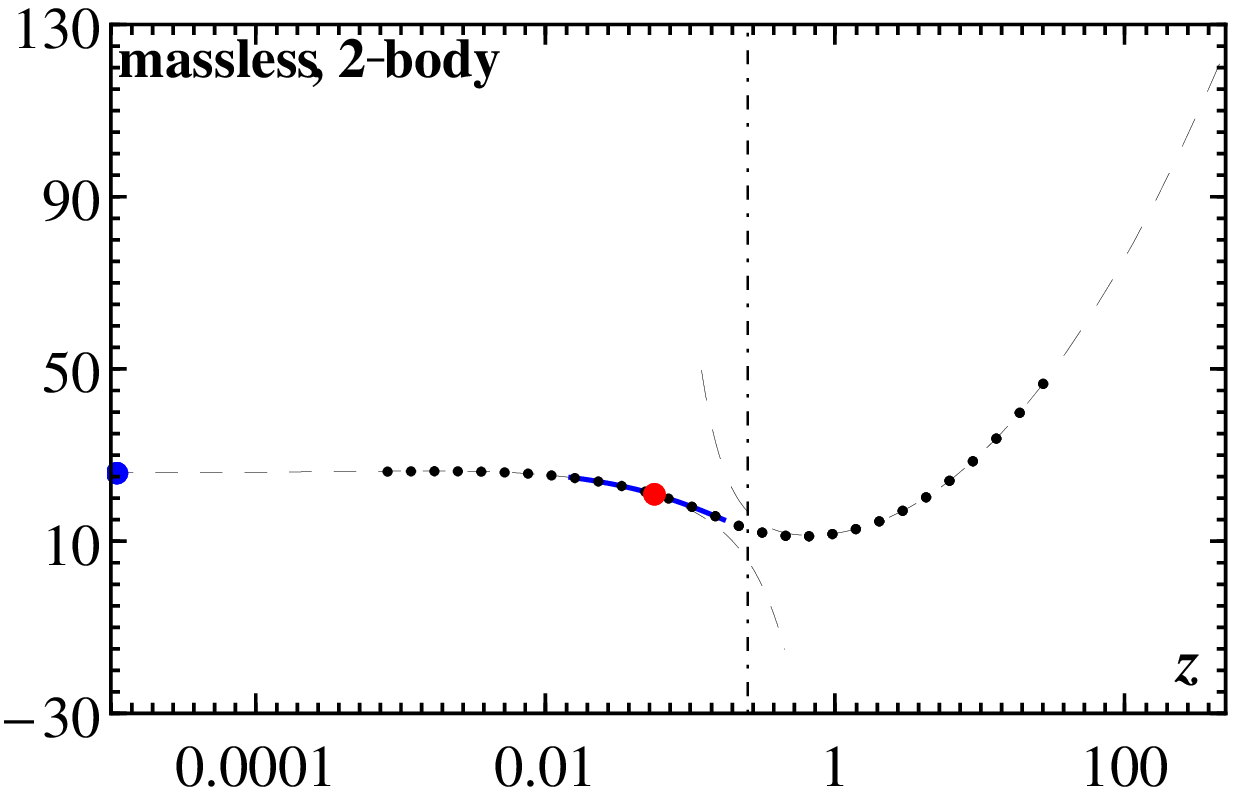} &\hspace{0.3cm}&
\includegraphics[scale=0.53]{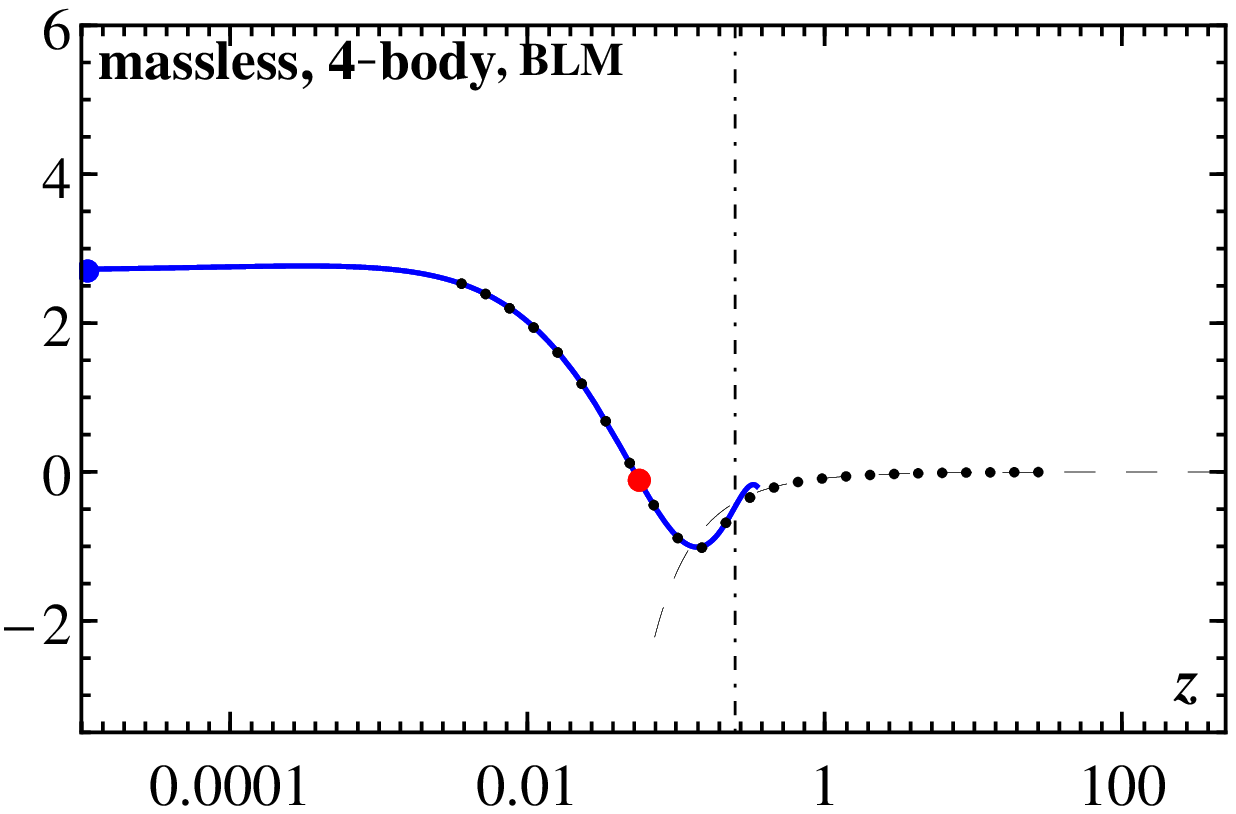}\\
\multicolumn{3}{c}{\includegraphics[scale=0.53]{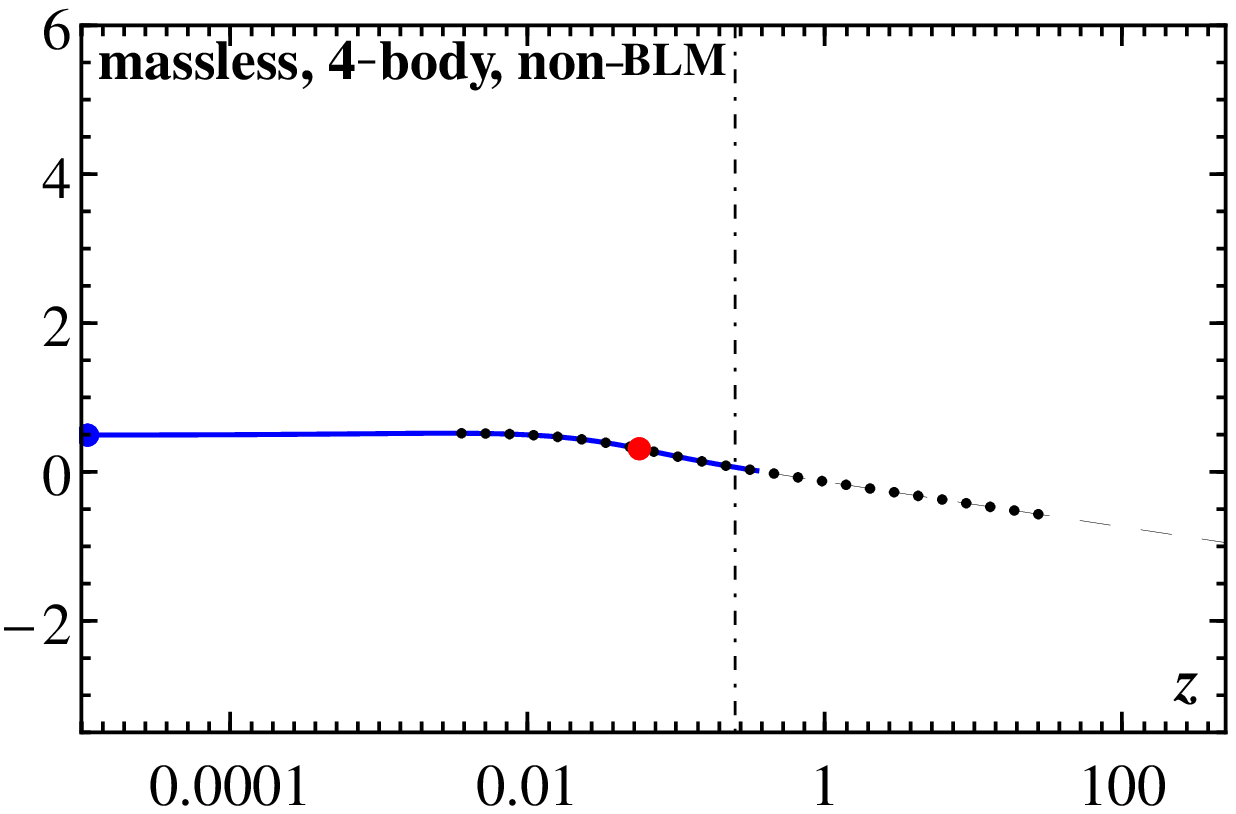}}
\end{tabular}
\caption {\sf Contributions to $\G_{27}^{(2)}$ from diagrams with closed loops of massless fermions - see the text.
They have already been multiplied by $n_l=3$, i.e. the number of flavours we treat as massless.}
\label{fig:z-plots1}
\end{center}
\end{figure}

After setting the renormalization scale squared to $\mu_b^2 = e^{\gamma}m_b^2/(4\pi)$
(with $\gamma$ being the Euler-Mascheroni constant), the MIs are
multiplied by appropriate powers of $m_b$, to make them
dimensionless. They depend on two parameters only: the dimensional
regularization parameter $\ep$, and the quark mass ratio $z =
m_c^2/m_b^2$. In each family separately, the MIs $M_k(z,\ep)$ satisfy
the Differential Equations (DEs)
\be
\f{d}{dz} M_k(z,\ep) = \sum_l R_{kl}(z,\ep) M_l(z,\ep),
\ee
where the rational functions $R_{kl}(z,\ep)$ on the r.h.s.\ are determined~\cite{Kotikov:1990kg,Remiddi:1997ny,Gehrmann:1999as} from the IBP, too.\footnote{
Getting a closed system of such DEs usually requires including several new MIs w.r.t.\ those entering the expression for $\G_{27}^{(2)}$.}
Similar equations are explicitly displayed in Eq.~(3.6) of Ref.~\cite{Misiak:2017woa} where ultraviolet counterterm contributions to $\G_{27}^{(2)}$ have been determined.

We solve the DEs using the same method as in Refs.~\cite{Boughezal:2007ny,Misiak:2017woa,ARthesis}.
The MIs are expanded in $\ep$ to appropriate powers, with the expansion coefficients being functions of $z$ only.
Boundary conditions for these functions at large $z$ are found using asymptotic expansions~\cite{Smirnov:2002pj}.
Next, the variable $z$ is treated as complex, and the DEs are
numerically solved along half-ellipses in the $z$-plane, to bypass
singularities on the real axis.

In practice, the codes {\tt q2e} and {\tt exp}~\cite{Harlander:1997zb,Seidensticker:1999bb}
are used to determine the asymptotic expansions at large $z$. Coefficients at
subsequent powers of $1/z$ are given in terms of one-, two- and
three-loop single-scale integrals, either massive tadpoles or
propagator-type ones with unitarity cuts (see Fig.~\ref{fig:SSIs}). Only at the
level of the latter integrals, we perform cross-family identification,
which gives us ${\mathcal O}(50)$ essentially different and
non-vanishing integrals. They are evaluated~\cite{SSI} using
various techniques, in particular the Mellin-Barnes one. Once
the large-$z$ expansions are found, numerical solutions of the DEs
starting from the boundary at $z=20$ are worked out using the code
{\tt ZVODE}~\cite{zvode} upgraded to quadrupole-double precision
with the help of the {\tt QD}~\cite{qd} computation
package. Half-ellipses of various sizes are considered to test the
numerical stability.
\begin{figure}[t]
\begin{center}
\begin{tabular}{ccc}
\includegraphics[scale=0.53]{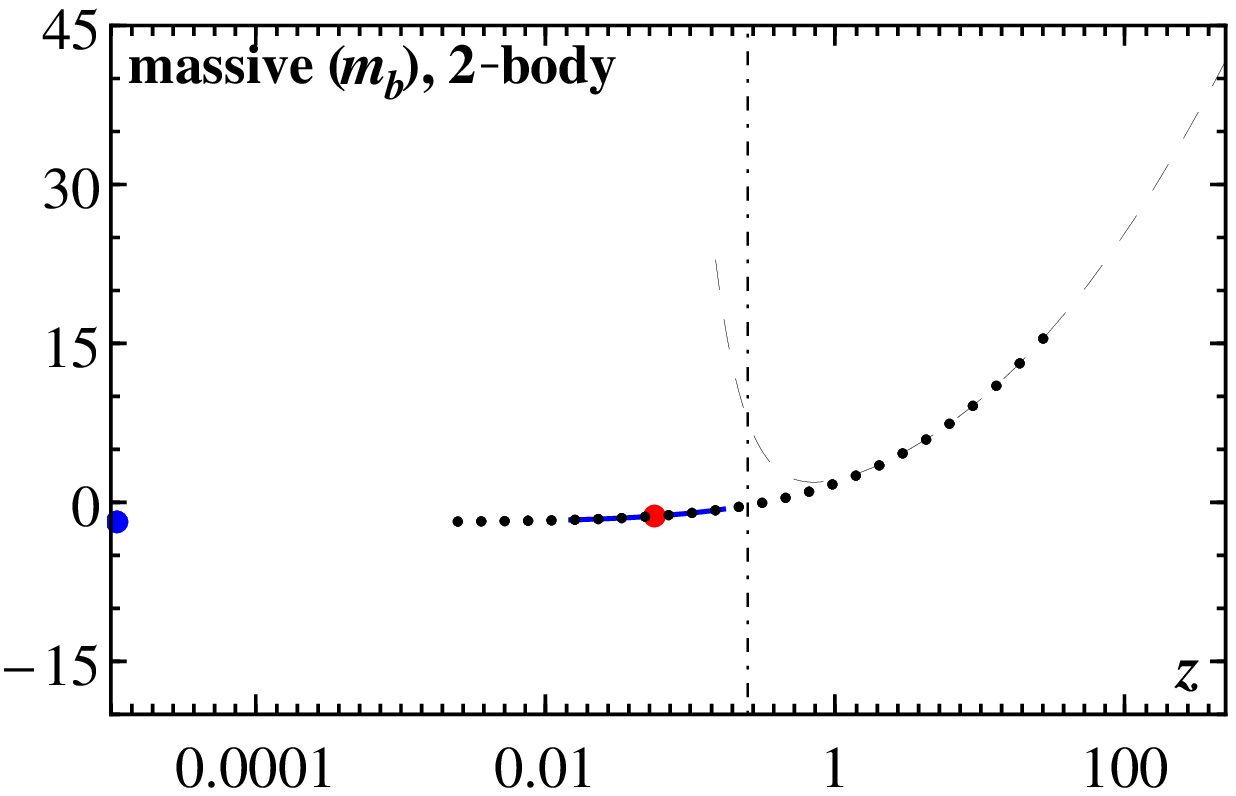} &\hspace{0.3cm}&
\includegraphics[scale=0.52]{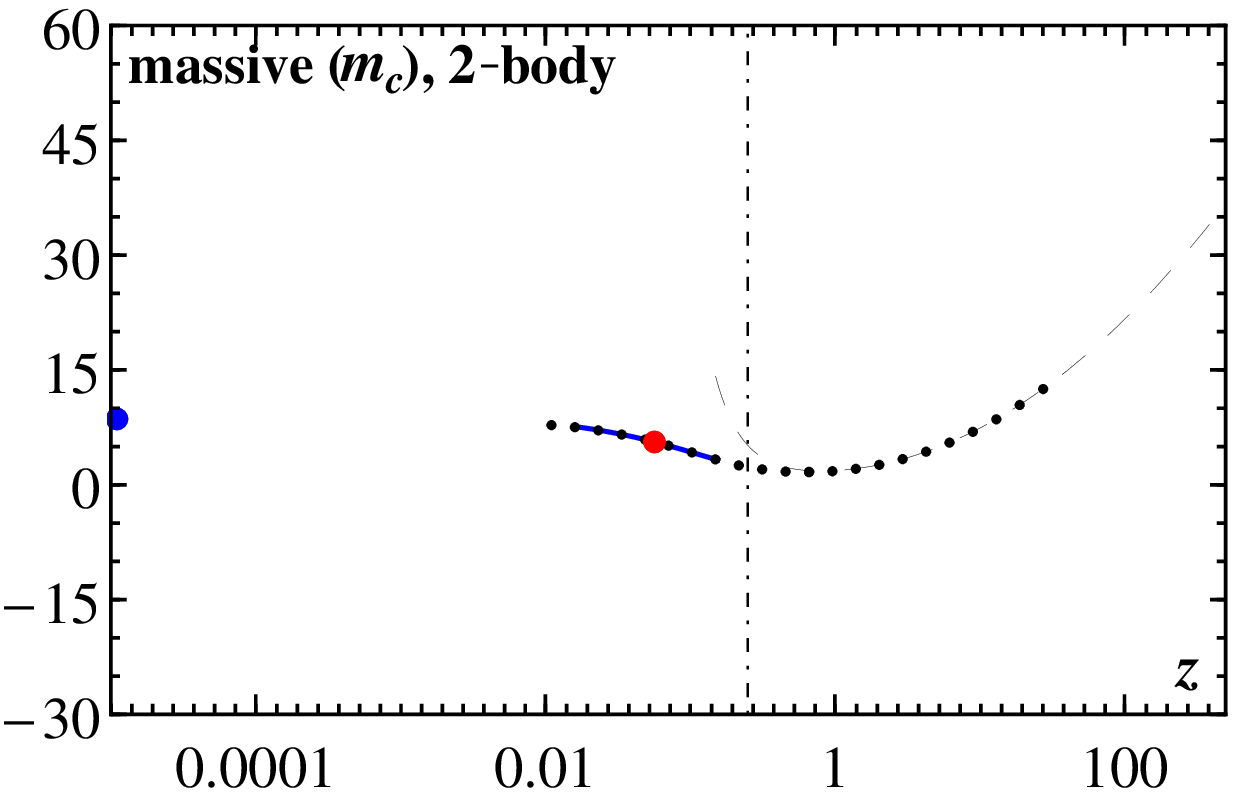}
\end{tabular}
\caption {\sf Contributions to $\G_{27}^{(2)}$ from diagrams with closed loops of massive fermions - see the text.}
\label{fig:z-plots2}
\end{center}
\end{figure}

At present, our IBP reduction for the full $\G_{27}^{(2)}$ is (almost)
completed, and the evaluation of the boundary conditions is being
finalized~\cite{SSI}. However, for the diagrams with closed fermionic
loops (as the ones in the first row of
Fig.~\ref{fig:sample.diagrams}), the DEs are already solved, and we
are ready to present the final results. They are plotted in Figs.~\ref{fig:z-plots1}
and~\ref{fig:z-plots2} as functions of $z$.

The displayed results correspond to various contributions to $\G_{27}^{(2)}$ 
renormalized in the $\overline{\rm MS}$ scheme with $\mu_b^2 = m_b^2$
(or, equivalently, in the ${\rm MS}$ scheme with $\mu_b^2 = e^{\gamma}m_b^2/(4\pi)$).
The renormalization has been performed with the help of the counterterm contributions
evaluated\footnote{
In the charm loop case (the right plot in Fig.~\ref{fig:z-plots2}), we had to rely on our so-far unpublished results for
the UV counterterms -- see the Appendix.}
in Refs.~\cite{Misiak:2017woa,ARthesis}. In all the plots, the black
dots correspond to numerical solutions that we have obtained using the
DEs. Dots corresponding to the physical value of $z$ are bigger and
highlighted in red.  Blue dots of similar size on the left boundaries
of each plot indicate the $z\to 0$ limits for each contribution, known
from the calculation in Ref.~\cite{Czakon:2015exa}. Thin dashed curves
continuing to large values of $z$ describe our large-$z$ expansions
evaluated up to ${\mathcal O}(1/z^2)$ (see the
Appendix). The dash-dotted vertical lines indicate the $c\bar c$
production threshold at $z=1/4$, in the vicinity of which neither the
large-$z$ nor the small-$z$ expansions are expected to converge well.

In Fig.~\ref{fig:z-plots1}, three distinct contributions from diagrams
with closed massless fermion loops are presented. The first (upper
left) plot corresponds to diagrams with two-body cuts. The thin dashed
line in the small-$z$ region shows the analytic expansion in powers of
$z$ evaluated in Ref.~\cite{Bieri:2003ue}. It is the only case for
which such an expansion is known. The solid blue curve shows the
numerical fit corresponding to Eq.~(3.2) of
Ref.~\cite{Boughezal:2007ny} where a numerical method (identical to
ours) has been used.
\begin{figure}[t]
\begin{center}
\begin{tabular}{ccc}
\includegraphics[width=5cm,angle=0]{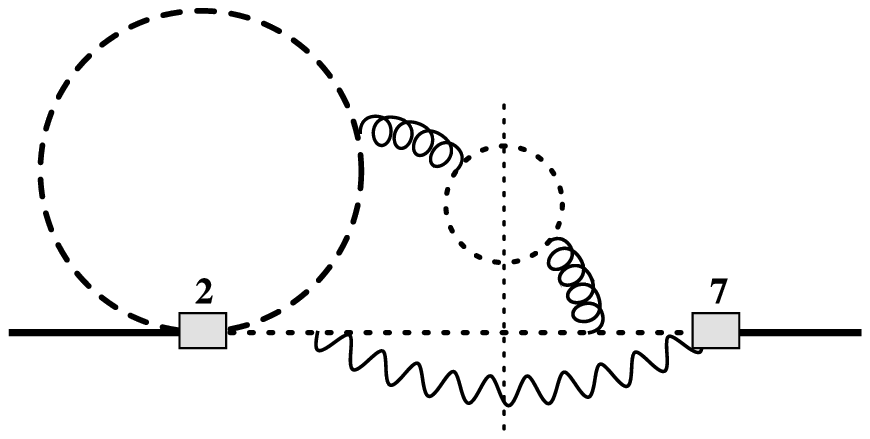} &~~~&
\includegraphics[width=5cm,angle=0]{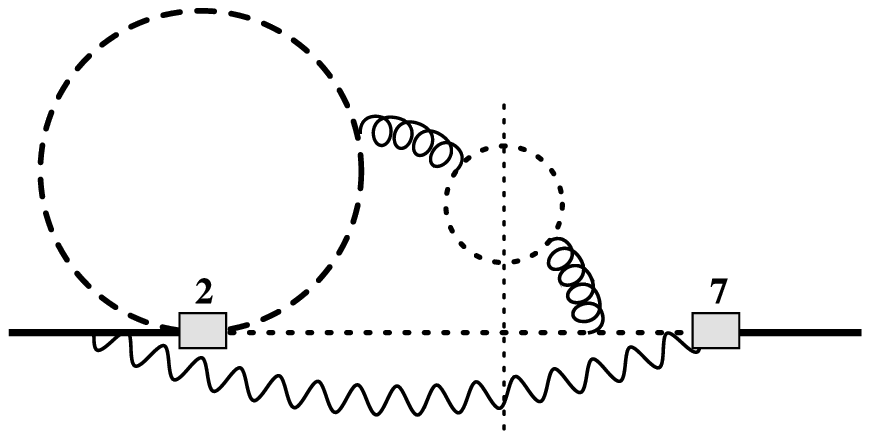} \\
\includegraphics[width=5cm,angle=0]{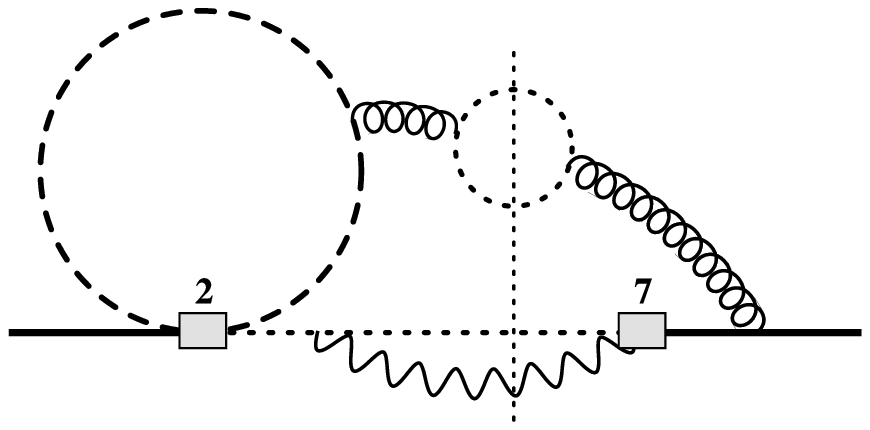} &&
\includegraphics[width=5cm,angle=0]{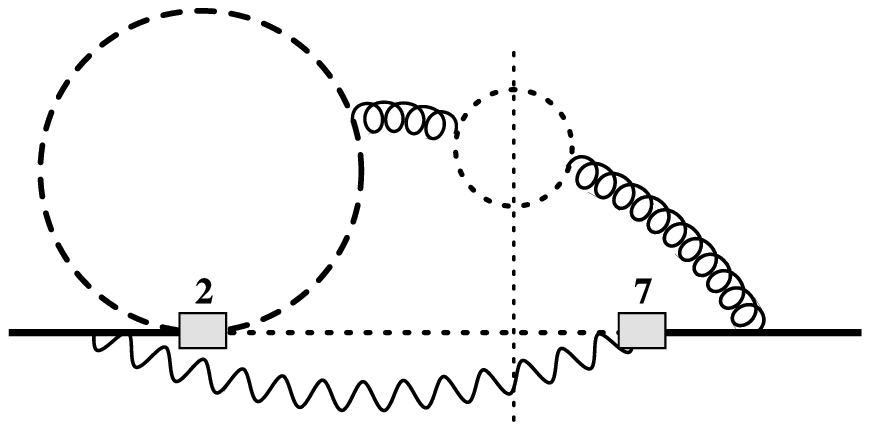} 
\end{tabular}
\end{center}
\caption{\sf Diagrams corresponding to the last (lower) plot in Fig.~\ref{fig:z-plots1}.} \label{fig:kappa.diags}
\end{figure}

The second (upper right) plot of Fig.~\ref{fig:z-plots1} shows all the
four-body-cut contributions except the diagrams displayed in
Fig.~\ref{fig:kappa.diags}. The latter diagrams have been skipped\footnote{
Arguments in favour of not including them in the BLM approach can be
found below Eq.~(12) of Ref.~[24]. They are correlated via
renormalization group with tree-level $b \to s q\bar q\gamma$ matrix
elements of the penguin four-quark operators.}
in evaluating the photon spectrum in the Brodsky-Lepage-Mackenzie
(BLM)~\cite{Brodsky:1982gc} approximation by the authors of
Refs.~\cite{Ligeti:1999ea,Misiak:2010tk}. The solid blue curve is
based on the numerical fit from Eq.~(3.6) of
Ref.~\cite{Czakon:2015exa} that corresponds to no restriction on
$E_\gamma$, and has been obtained as a by-product of the calculation
in Ref.~\cite{Misiak:2010tk}.

The third (bottom) plot in Fig.~\ref{fig:z-plots1} corresponds to the
very diagrams from Fig.~\ref{fig:kappa.diags}. In this case, no
numerical result valid for arbitrary $m_c$ has existed prior to our
present calculation. For $z < \f14$, we can describe our findings by
the following fit:
\be \label{kappa.fit}
\Delta^{\mbox{\tiny 4-b \cancel{BLM}}}_{\scs m=0} \G_{27}^{(2)} = 3 \left[ 0.164 + 0.13\, z^\f12 - 21.51\, z +
    68.10\, z^\f32 - 46.12\, z^2 + (-3.23\, z  + 18.23\, z^2) \ln z\right].~
\ee
It is shown as a solid blue curve in the considered plot. A quick look
at Fig.~\ref{fig:kappa.diags} is sufficient to realize that
$\Delta^{\mbox{\tiny 4-b \cancel{BLM}}}_{\scs m=0} \G_{17}^{(2)} = -\f16 \Delta^{\mbox{\tiny 4-b \cancel{BLM}}}_{\scs m=0} \G_{27}^{(2)}$,
due to the identity $T^a T^b T^a = -\f16 T^b$ for the $SU(3)_c$
generators. The same relative colour factor is valid for all the plots
in Figs.~\ref{fig:z-plots1} and~\ref{fig:z-plots2}.

Fig.~\ref{fig:z-plots2} shows contributions to $\G_{27}^{(2)}$ from
diagrams with closed loops of quarks with masses $m_b$ (left) and
$m_c$ (right). Only the two-body cuts are included. The solid
blue lines correspond to the numerical fits from Eqs.~(3.3) and (3.4) of
Ref.~\cite{Boughezal:2007ny}.  In these cases, no four-body
cuts are allowed, as the state $X^p_s$ in Eq.~(\ref{def.ghat}) is
assumed to contain no charm quarks. We do not consider three-body
cuts here, as their effect can be included by multiplying the
well-known three-body contribution to $\G_{27}^{(1)}$ by finite
coefficients originating from\footnote{
$Z_G^{\rm OS}$ stands for the on-shell renormalization constant of the
gluon wave-function, while $Z_g$ renormalizes the QCD gauge coupling
in the MS scheme.}
%
%
$\;Z_G^{\rm OS} Z_g^2-1$. The corresponding term in Eq.~(3.8) of
Ref.~\cite{Czakon:2015exa} comes at the end of the first line of the
expression for $K_{27}^{(2)}$.

As evident from the plots, our results are in perfect agreement with
all the previously available expansions and fits. It is particularly
important in the massive case (Fig.~\ref{fig:z-plots2}) where our
verification comes as the first one from an independent group. Let
us note that the contribution displayed in the right plot of
Fig.~\ref{fig:z-plots2} affects ${\mathcal B}_{s \gamma}$ by around
$-2.1\%$, which should be compared to the current ($\pm 4.5\%$) and
expected future ($\pm 2.6\%$) experimental accuracies mentioned in
Sec.~\ref{sec:intro}. The massless results from the upper two plots
of Fig.~\ref{fig:z-plots1} have already been cross-checked before.

As far as the new contribution (the third plot in
Fig.~\ref{fig:z-plots1}) is concerned, it has so far been included in
the interpolated part of the NNLO correction, and resulted in a tiny
effect, around one per-mille of the decay rate only. Now we remove it
from the interpolated part and replace by the fit in
Eq.~(\ref{kappa.fit}). It turns out that the interpolation estimate was
correct within ~$\sim\!\! 10\%$ of the considered contribution, so the
effect remains tiny.

\section{\boldmath Updated SM predictions for ${\mathcal B}_{s \gamma}$ and $R_\gamma$ \label{sec:num}}

In the present section, we work out updated SM predictions for ${\mathcal B}_{s \gamma}$, as well as
for the ratio
$R_\gamma \equiv {\mathcal B}_{(s+d) \gamma}/{\mathcal B}_{c\ell\bar\nu}$,
where ${\mathcal B}_{c\ell\bar\nu}$ is the CP- and isospin-averaged
branching ratio of the inclusive semileptonic decay.  Our main
motivation for performing an update right now is not due to the NNLO
corrections evaluated in the previous section. The new contribution is
tiny, while the sizeable ones (that we have confirmed) were
already included in the phenomenological analysis of
Ref.~\cite{Czakon:2015exa}. However, there has been an important progress in
estimating non-perturbative effects (see below). An
update of the SM prediction should thus be performed right now, even
though the $m_c$-interpolation uncertainty remains essentially
unchanged.

The first improvement in estimating the non-perturbative effects
becomes possible thanks to the new Belle measurement of the isospin asymmetry
\be \label{Belle.isosp}
\Delta_{0-} \equiv
\f{\Gamma[\bar B^0 \to X_s\gamma]-\Gamma[B^- \to X_s\gamma]}{\Gamma[\bar B^0 \to X_s\gamma]+\Gamma[B^- \to X_s\gamma]} 
= (-0.48 \pm 1.49 \pm 0.97 \pm 1.15 )\%~\mbox{\cite{Watanuki:2018xxg}}.
\ee
In the SM, the dominant contribution to this asymmetry arises from a
process where no hard photon but rather a hard\footnote{
with momentum of order $m_b$ but possibly much smaller virtuality}
gluon is emitted in the $b$-quark decay~\cite{Lee:2006wn}. Next, the
gluon scatters on the valence quark, which results in emission of a
hard photon. Instead of the valence quark, also a sea quark ($u$, $d$
or $s$) can participate in such a Compton-like scattering. Taking this
fact into account, one can write the decay rates as
\bea
\Gamma[     B^{-\!} \to X_s\gamma] &\simeq& A + B Q_u + C Q_d + DQ_s,\nnb\\[1mm]
\Gamma[\bar B^{0\,} \to X_s\gamma] &\simeq& A + B Q_d + C Q_u + DQ_s,\label{def.ABCD}
\eea
where $Q_{u,d,s}$ denote electric charges of the quarks participating
in the Compton-like scattering, while the quantities $A$, \ldots, $D$
are given by interferences of various quantum amplitudes whose
explicit form is inessential here. Since the considered effect
gives only a small correction to the decay rate ($B,C,D \ll
A$), quadratic terms in $Q_{u,d,s}$ have been neglected above. We have
also neglected isospin violation in the quark masses ($m_u\neq m_d$)
and in the electromagnetic corrections to the $\bar B$-meson wave
functions (suppressed by extra powers of $\alpha_{em}$).

The leading term $A$ contains the dominant contribution originating
from the operator $Q_7$. The corrections $B$, $C$, $D$ are suppressed
w.r.t.\ $A$ both by $g_s^2$ (as the gluon is hard) and by $\Lambda/m_b$,
with $\Lambda \sim \Lambda_{\rm QCD}$. The latter suppression can be
intuitively understood by realizing that the gluon scatters on
remnants of the $\bar B$ meson, i.e.\ on a diluted target whose size
scales like $1/\Lambda$. Such a suppression is confirmed in
Refs.~\cite{Lee:2006wn,Benzke:2010js} where the Soft-Collinear
Effective Theory (SCET) has been applied to analyze non-perturbative
corrections to ${\mathcal B}_{s \gamma}$.

From Eq.~(\ref{def.ABCD}), one easily obtains the isospin-averaged decay rate
\be
\Gamma \simeq A + \f12 (B+C)(Q_u + Q_d) + DQ_s \equiv A + \delta\Gamma_c, 
\ee
and the isospin asymmetry
\be
\Delta_{0-} \simeq \f{C-B}{2\Gamma} (Q_u-Q_d).
\ee
It follows that the relative correction to the isospin-averaged decay rate that arises due to the considered effect reads
\be \label{su3f}
\f{\delta\Gamma_c}{\Gamma} \simeq \f{(B+C)(Q_u+Q_d) + 2 D Q_s}{(C-B)(Q_u-Q_d)} \Delta_{0-} = \f{Q_u+Q_d}{Q_d-Q_u} \left[ 1 + 2\,\f{D-C}{C-B}\right] \Delta_{0-},
\ee
where, in the last step, $Q_s = -Q_u - Q_d$ has been used. The second term
in the square bracket vanishes in the $SU(3)_F$ limit, i.e.\ when the
three lightest quarks are treated as mass-degenerate. In this limit,
as observed in Ref.~\cite{Misiak:2009nr},~ $\delta\Gamma_c/\Gamma$~
and~ $\Delta_{0-}$~ are related to each other in a simple manner that is
free from non-perturbative uncertainties.  The authors of
Ref.~\cite{Benzke:2010js} suggested $\pm 30\%$ as an uncertainty estimate
stemming from the $SU(3)_F$-violating effect in Eq.~(\ref{su3f}). Following this
suggestion, we find
\be \label{num78}
\f{\delta\Gamma_c}{\Gamma} = -\f13 ( 1 \pm 0.3) \Delta_{0-} = (0.16 \pm 0.74)\%,
\ee
where the experimental errors from Eq.~(\ref{Belle.isosp}) were
combined in quadrature, giving $\Delta_{0-} = (-0.48 \pm 2.12)\%$; next,
the multiplicative factor was taken into account as follows~\cite{Goodman:1960var}:
\be
( 1 \pm 0.3) (-0.48 \pm 2.12)\% = \left(-0.48 \pm \sqrt{ 2.12^2 + (0.3 \cdot 0.48)^2 + (0.3 \cdot 2.12)^2}\right)\%.
\ee

In the above considerations, we have treated the measured
$\Delta_{0-}$ in Eq.~(\ref{Belle.isosp}) as already extrapolated from
the experimental cutoff of $E_0 = 1.9\,$GeV down to our default $E_0 =
1.6\,$GeV, even though no such extrapolation has actually been done in
Ref.~\cite{Watanuki:2018xxg}, i.e. Eq.~(\ref{Belle.isosp}) corresponds
to $E_0 = 1.9\,$GeV. A devoted analysis would be necessary to estimate
the extrapolation effects in this case. However, we expect such
effects to be negligible w.r.t.\ the experimental uncertainties in
Eq.~(\ref{Belle.isosp}).

If the uncertainty on the r.h.s.\ of Eq.~(\ref{num78}) is treated as $1\sigma$ of a
Gaussian distribution, then the 95\%$\,$C.L. range is $[-1.3,+1.6]\%$. The
corresponding\footnote{
Our $\delta\Gamma_c/\Gamma$ and their ${\mathcal F}^{\rm exp}_{78}$ are estimated in a similar way.}
range $[-1.4,+2.0]\%$ in Sec.~3.5 of Ref.~\cite{Gunawardana:2019gep}
is somewhat wider due to a different method of combining uncertainties
and using the PDG~\cite{Tanabashi:2018oca,PDGupdate} central value of~ $-0.6\%$~ for $\Delta_{0-}$.
When determining our SM predictions below, we calculate ${\mathcal B}_{s \gamma}$
without including the photon emission from the valence/sea
quarks and, in the final step, we multiply by $\left(1 +
\f{\delta\Gamma_c}{\Gamma}\right)$, employing the number from the r.h.s.\
of Eq.~(\ref{num78}).

Another important non-perturbative correction to be considered arises
in the interference of $Q_{1,2}$ and $Q_7$. Its presence in the
inclusive $\bar B \to X_s \gamma$ rate was first pointed out in
Ref.~\cite{Voloshin:1996gw}. It amounts to around $+3\%$ of ${\mathcal
B}_{s \gamma}$, as established in Ref.~\cite{Buchalla:1997ky} at the
leading order of an expansion in powers of~ $m_b\Lambda/m_c^2$.~ The
corresponding leading contribution to $N(E_0)$ in Eq.~(\ref{brB})
reads
\be \label{Volcor}
\delta N_V = -\f{\mu_G^2}{27m_c^2} C_7(\mu_b) \left( C_2(\mu_b) -\f16 C_1(\mu_b) \right),
\ee
where $\mu_G^2 \simeq 0.3\,{\rm GeV}^2$ is one of the HQET parameters
that matter in the determination of $C$ in Eq.~(\ref{phase}). Since~
$m_b\Lambda/m_c^2$~ is not a small parameter, the authors of
Ref.~\cite{Benzke:2010js} argued that no expansion in its powers can
be used at all. Instead, they estimated the considered correction in
the framework of SCET, where essential constraints on models of the
relevant soft function came from moments of the semileptonic $\bar B
\to X_c \ell \bar\nu$ decay spectra. A recent update of these
estimates in Ref.~\cite{Gunawardana:2019gep} implies that $\delta N_V$
(\ref{Volcor}) needs to be multiplied by
\be \label{kappaV}
\kappa_V = 1 - \f{27 m_c^2\Lambda_{17}}{m_b\mu_G^2} = 1.2 \pm 0.3.
\ee
The final numerical value above has been derived by us from ranges for
$\Lambda_{17}$ given in Ref.~\cite{Gunawardana:2019gep}, assuming that
these ranges can be interpreted as $1\sigma$ ones. The remaining
parameters on which $\kappa_V$ depends were set to the values
corresponding to the widest range for $\Lambda_{17}$ in
Ref.~\cite{Gunawardana:2019gep}.

Since the expression for $\delta N_V$ (\ref{Volcor}) is calculated at
the leading order in QCD only, the renormalization scheme for $m_c^2$
in the denominator is unspecified. We assume that the corresponding
uncertainty is included in the overall~ $\pm 3\%$~ higher-order one
that is being retained the same as in Ref.~\cite{Czakon:2015exa}. As
the total effect from $\delta N_V$ amounts to around $3\%$ in
${\mathcal B}_{s \gamma}$, uncertainties due to scheme-dependence of
$m_c$ in $\delta N_V$ can safely be treated this way. In our numerical
calculations, the quark masses and HQET parameters are included with
a full correlation matrix (see Appendix D of
Ref.~\cite{Czakon:2015exa}), except for the very $m_c$ in $\delta N_V$
that is now fixed to $1.17\,$GeV. The parameter $\kappa_V$
(\ref{kappaV}) will be treated as uncorrelated.

Apart from the two effects we have discussed above, the authors of
Ref.~\cite{Benzke:2010js} identified a third source of uncertain
contributions to $N(E_0)$ that arise at the order ${\mathcal
O}(\Lambda/m_b)$. They come proportional to $|C_8(\mu_b)|^2$, where
$C_8$ is the Wilson coefficient of the gluonic dipole operator
\be \label{Q8}
Q_8 = \fm{g_s m_b}{16\pi^2} (\bar{s}_L \sigma^{\mu \nu} T^a b_R) G^a_{\mu \nu}.
\ee
Previous estimates of these corrections in
Refs.~\cite{Kapustin:1995fk,Ferroglia:2010xe} focused on large
collinear logarithms $\ln\f{m_b}{m_s}$~ that are present in the corresponding
contributions to $P(E_0)$. In Ref.~\cite{Czakon:2015exa}, such logarithms were varied in the range
$\left[ \ln 10, \ln 50 \right] \simeq \left[ \ln\f{m_B}{m_K}, \ln\f{m_B}{m_\pi} \right]$,
which served as a crude estimate of the very uncertain but otherwise
small contributions to ${\mathcal B}_{s \gamma}$ where light hadron
masses are the physical collinear regulators. However,
according to Ref.~\cite{Benzke:2010js}, possible non-perturbative
effects that come multiplied by $|C_8(\mu_b)|^2$ can be unrelated to
collinear logarithms, and affect ${\mathcal B}_{s \gamma}$ by relative
corrections in the range $[-0.3, 1.9]\%$ with respect to the
$\f{m_b}{m_s}=50$ case, for $\mu_b = 1.5\,$GeV and $E_0 =
1.6\,$GeV. Numerically, we can reproduce this range by performing a
replacement
\be \label{kappa88}
\ln\f{m_b}{m_s} \to \kappa_{88} \ln 50 \hspace{1cm} \mbox{with} \hspace{1cm}  
\kappa_{88} = 1.7 \pm 1.1
\ee
in all the perturbative contributions proportional to $|C_8(\mu_b)|^2$. 

In the following, we shall treat the quantities 
$\f{\delta\Gamma_c}{\Gamma}$~(\ref{num78}), $\kappa_V$~(\ref{kappaV}) and $\kappa_{88}$~(\ref{kappa88})
on equal footing with all the other parameters that ${\mathcal B}_{s \gamma}$
depends on. Since they account for all the non-perturbative effects
estimated in Refs.~\cite{Gunawardana:2019gep,Benzke:2010js}, we shall
no longer include the overall $\pm 5\%$ non-perturbative uncertainty that
entered the analysis of Ref.~\cite{Czakon:2015exa} as an input from
Ref.~\cite{Benzke:2010js}. This way we determine our updated SM predictions
for ${\mathcal B}_{s \gamma}$ and $R_\gamma$ in the SM, namely
\be \label{update.main}
{\mathcal B}_{s \gamma} = (3.40 \pm 0.17) \times 10^{-4} 
\hspace{1cm} \mbox{and} \hspace{1cm}  
R_\gamma = (3.35 \pm 0.16) \times 10^{-3},
\ee
for $E_0 = 1.6\,$GeV. The overall uncertainties have been obtained by
combining in quadrature the ones stemming from higher-order effects
($\pm 3\%$), interpolation in $m_c$ ($\pm 3\%$), as well as the
parametric uncertainty where all the non-perturbative ones are now contained.
Not only $\f{\delta\Gamma_c}{\Gamma}$, $\kappa_V$ and $\kappa_{88}$
but several other inputs parameterize non-perturbative effects,
too, namely the collinear regulators (see above), as well as the HQET
parameters that enter either directly or via the semileptonic
phase-space factor $C$ (\ref{phase}). In the ${\mathcal B}_{s \gamma}$ case,
our parametric uncertainty amounts to $\pm 2.5\%$ at present. All the input
parameters listed in Appendix~D of Ref.~\cite{Czakon:2015exa}
have been retained unchanged.

The overall uncertainty in $R_\gamma$ (\ref{update.main}) amounts to
$\pm 4.8\%$, noticeably improved w.r.t.\ to $\pm 6.7\%$ in
Ref.~\cite{Misiak:2015xwa}. The main reason for the improvement comes
from the updated estimate in Ref.~\cite{Gunawardana:2019gep} of
the non-perturbative uncertainty that stems from $\Lambda_{17}$ in
Eq.~(\ref{kappaV}). Further improvement requires removing the
$m_c$-interpolation, and re-considering the higher-order and
parametric uncertainties. If they remain unchanged, the expected
future accuracy in the SM prediction for ${\mathcal B}_{s \gamma}$
amounts to $\sqrt{3^2 + 2.5^2}\,\% \simeq 3.9\%$, still somewhat
behind the experimental expectation of $\pm 2.6\%$ that was mentioned
above Eq.~(\ref{brB}).

In many BSM theories, extra additive contributions $\Delta
C_{7,8}$ to the Wilson coefficients of the operators $Q_7$~(\ref{ops})
and $Q_8$~(\ref{Q8}) at the electroweak matching scale $\mu_0$ are the
only relevant reason for shifting ${\mathcal B}_{s \gamma}$ and
$R_\gamma$ away from the SM predictions. So long as no accidental
cancellations occur, effects due to $\Delta C_{7,8}$ must be small
whenever the current experimental constraints are satisfied. At such
points in the BSM parameter spaces, ${\mathcal B}_{s \gamma}$ and
$R_\gamma$ can accurately be calculated from the following simple
linearized expressions
\bea 
{\mathcal B}_{s \gamma} \times 10^4 &=& (3.40 \pm 0.17) - 8.25\,\Delta C_7 - 2.10\, \Delta C_8\,,\nnb\\[1mm]
R_\gamma \times 10^3 &=& (3.35 \pm 0.16) - 8.08\,\Delta C_7 - 2.06\, \Delta C_8\,,\label{bsm}
\eea
where $\mu_0=160\,$GeV has been chosen. The above equations are updates
of similar ones in Eq.~(10) of Ref.~\cite{Misiak:2015xwa}. Analytic formulae
for the Wilson coefficients at $\mu_0$ in a wide class of BSM theories
can be found in Ref.~\cite{Bobeth:1999ww}.

In the specific case of the Two-Higgs-Doublet Model, Eq.~(\ref{bsm})
can be replaced by expressions including all the NLO and NNLO QCD
matching corrections~\cite{Hermann:2012fc}. The resulting $95\%\,$C.L.\
lower bound from $R_\gamma$ on the charged Higgs boson mass in
Model-II, evaluated along the same lines\footnote{
The corresponding bound in the conclusions of Ref.~\cite{Misiak:2017bgg} amounted to $580\,$GeV.}
as in Ref.~\cite{Misiak:2017bgg}, yields $800\,$GeV.

\section{Summary \label{sec:summary}}

We reported on our calculation of the NNLO QCD corrections to
${\mathcal B}_{s \gamma}$ without interpolation in $m_c$, and
presented final results for contributions originating from propagator
diagrams with closed fermion loops on the gluon lines. They correspond
either to the two-body ($s\gamma$) or four-body ($sq\bar q\gamma$)
final states. In all the previously investigated cases, we confirmed
the results from the literature, some of which had been obtained by a
single group only. The new part comes from four diagrams with
four-particle cuts that had not been determined before, as they are
not included in the BLM approximation. Their contribution turns out to
be tiny ($\sim 0.1\%$ of the decay rate) and quite well reproduced by
our former interpolation algorithm.

In view of the recent progress in estimating the non-perturbative
contributions, we performed an update of the phenomenological analysis
within the SM. The obtained results yield~
${\mathcal B}_{s \gamma} = (3.40 \pm 0.17) \times 10^{-4}$~
and~
$R_\gamma \equiv {\mathcal B}_{(s+d) \gamma}/{\mathcal B}_{c\ell\bar\nu} = (3.35 \pm 0.16) \times 10^{-3}$~
for~ $E_0 = 1.6\,$GeV. The main improvement in the uncertainty came
from the analysis in Ref.~\cite{Gunawardana:2019gep} where
non-perturbative effects in the $Q_{1,2}$-$Q_7$ interference were
re-analyzed.

The next contribution to suppressing the overall theoretical
uncertainty is expected from the calculation of $\G_{17}^{(2)}$ and
$\G_{27}^{(2)}$ for $E_0=0$ and at the physical value of $m_c$, thereby
removing the need for $m_c$-interpolation in these quantities.

\section*{Acknowledgments}

We are grateful to Alexander Smirnov and Johann Usovitsch for
providing help to us as users of {\tt FIRE} and {\tt KIRA},
respectively. We would like to thank Gil Paz for extensive discussions
concerning the non-perturbative contributions. The research of AR and
MS has been supported by the Deutsche Forschungsgemeinschaft (DFG,
German Research Foundation) under grant 396021762 --- TRR 257
``Particle Physics Phenomenology after the Higgs Discovery''. MM has
been partially supported by the National Science Center, Poland, under
the research project 2017/25/B/ST2/00191, and the HARMONIA project
under contract UMO-2015/18/M/ST2/00518. This research was
supported in part by the PL-Grid Infrastructure.

\section*{Note added in the proofs}

While the present article was being reviewed for publication, a new
paper~\cite{Benzke:2020htm} on non-perturbative effects in the
$Q_{1,2}$-$Q_7$ interference appeared on the arXiv. To replace the
estimates of Ref.~\cite{Gunawardana:2019gep} by those of
Ref.~\cite{Benzke:2020htm} in our approach, one would need to use
$\kappa_V = 1.7 \pm 0.8$ in Eq.~(\ref{kappaV}). This would shift our
prediction for ${\mathcal B}_{s \gamma}$ from $(3.40 \pm 0.17) \times
10^{-4}$ to $(3.45 \pm 0.19) \times 10^{-4}$, and strengthen the
constraint on $M_{H^\pm}$ even more.
%
%
However, the extreme values of $\Lambda_{17}$ in
Ref.~\cite{Benzke:2020htm} originate from soft function models with
quite a rich structure. Such soft functions are related to
energy-momentum distributions of gluons inside the QCD ground states
($B$ mesons), in which case encountering large numbers of extrema and
zero points seems unlikely. Therefore, our preference is to retain
$\kappa_V$ as it stands in Eq.~(\ref{kappaV}) for evaluating the SM
predictions for ${\mathcal B}_{s \gamma}$ and $R_\gamma$.

\section*{Appendix: {\boldmath Large-$z$ expansions and
$\G_{47}^{(1)}$} with charm loops}
\def\theequation{A.\arabic{equation}}

In this appendix, we present large-$z$ expansions of the renormalized
contributions to $\G_{27}^{(2)}$ plotted in Figs.~\ref{fig:z-plots1}
and~\ref{fig:z-plots2}. They are shown by the thin dashed lines
reaching large values of $z$ in the corresponding plots. For the three
plots in Fig.~\ref{fig:z-plots1} that describe contributions from
diagrams with closed loops of massless fermions, the respective
expansions read\\[-3mm]
\bea
\Delta^{\mbox{\tiny 2-b}}_{\scs m=0} \G_{27}^{(2)} &=& 3 \left[ 
\fm{27650}{6561} + \fm{112}{243} L + \fm{8}{9} L^2 + \fm{1}{z} \left(\fm{10427}{30375} - \fm{8}{135} \pi^2 - \fm{572}{18225} L + \fm{38}{405} L^2 \right)\right.\nnb\\[2mm]
&+& \left. \fm{1}{z^2} \left( \fm{19899293}{125023500} - \fm{8}{405} \pi^2 - \fm{1628}{893025} L + \fm{86}{2835} L^2 \right)\right]
+ {\mathcal O}\left( \fm{1}{z^3} \right),\nnb\\[2mm]
\Delta^{\mbox{\tiny 4-b BLM}}_{\scs m=0}\, \G_{27}^{(2)} &=&
3 \left[ \fm{1}{z} \left( \fm{41}{108} - \fm{10}{243} \pi^2\right) + \fm{1}{z^2} \left( \fm{487}{3375} - \fm{2}{135} \pi^2 \right)\right]
+ {\mathcal O}\left( \fm{1}{z^3} \right),\nnb\\[2mm]
\Delta^{\mbox{\tiny 4-b \cancel{BLM}}}_{\scs m=0}\,  \G_{27}^{(2)} &=& 3 \left[-\fm{32}{729} (1+L) + \fm{1}{z} \left( -\fm{941}{7290} + \fm{16}{1215} \pi^2\right)
+ \fm{1}{z^2} \left(- \fm{10852}{212625} + \fm{44}{8505} \pi^2 \right)\right] + {\mathcal O}\left( \fm{1}{z^3} \right),\nnb\\ \ \\[-1cm]\nnb
\eea
where $L = \ln z$. The first expression above coincides with Eq.~(5.3) of Ref.~\cite{Misiak:2006ab}.

For the closed bottom loops (the left plot in Fig.~\ref{fig:z-plots2}), we find\\[-2mm]
\bea
\Delta^{\mbox{\tiny 2-b}}_{\scs m=m_b} \G_{27}^{(2)} &=&
 \fm{62210}{6561} + \fm{160}{729} \pi^2 - \fm{16 \pi}{9 \sqrt{3}} - 16 S_2
           + \left( \fm{464}{81} + \fm{160}{729} \pi^2 - \fm{16 \pi}{9 \sqrt{3}} - 16 S_2 \right) L + \fm{8}{9} L^2\nnb\\[2mm]
 &+& \fm{1}{z} \left(-\fm{30991}{10125} + \fm{656}{3645} \pi^2 + \fm{4 \pi}{45 \sqrt{3}} + \fm{64}{405} \zeta_3 + \fm{4}{5} S_2 - \fm{32972}{18225} L + \fm{38}{405} L^2\right)\nnb\\[2mm]
 && \hspace{-25mm} +~~
     \fm{1}{z^2} \left( -\fm{38874763}{25004700} - \fm{8}{1701} \pi^2 + \fm{26 \pi}{525 \sqrt{3}} + \fm{64}{2835} \zeta_3 + \fm{12}{35} S_2 - \fm{864896}{893025} L - \fm{418}{2835} L^2 \right)
 + {\mathcal O}\left( \fm{1}{z^3} \right), \label{mblarge}
\eea
where $S_2 = \f{4}{9 \sqrt{3}}\, {\rm Im}\!\left[ {\rm Li}_2 \left( e^{i\pi/3} \right)\right]$. Finally, for the closed charm loops (the right plot in Fig.~\ref{fig:z-plots2}),
the large-$z$ expansion reads
\bea
\Delta^{\mbox{\tiny 2-b}}_{\scs m=m_c} \G_{27}^{(2)} &=&
\fm{11018}{6561} + \fm{128}{243} L + \fm{200}{243} L^2
+ \fm{1}{z} \left( \fm{5714}{54675} + \fm{7}{81} \zeta_3 + \fm{2146}{18225} L + \fm{52}{405} L^2 \right)\nnb\\[2mm]
&+& \fm{1}{z^2} \left(-\fm{62075113}{428652000} + \fm{469}{5184} \zeta_3 - \fm{41987}{893025} L + \fm{92}{2835} L^2 \right)
+ {\mathcal O}\left( \fm{1}{z^3} \right). \label{mclarge}
\eea
Our results in Eqs.~(\ref{mblarge}) and (\ref{mclarge}) agree with the numerical ones in Eqs.~(A.1) and (A.2) of Ref.~\cite{Boughezal:2007ny}.
Analytical expressions for the leading terms agree with the findings of Ref.~\cite{Misiak:2010sk}.

Determining the renormalized results plotted in
Figs.~\ref{fig:z-plots1} and~\ref{fig:z-plots2} required taking into
account $\G_{47}^{(1){\rm bare}}$, i.e. three-loop counterterm diagrams with vertices proportional to
$Q_4 = (\bar{s}_L \gamma_{\mu} T^a b_L) \sum_q (\bar{q}\gamma^{\mu} T^a q)$.
An expression for this quantity in Eq.~(2.4) of
Ref.~\cite{Czakon:2015exa} contains no contributions from closed loops
of charm quarks, as all the other results in Sec.~2 of that
paper. Such contributions arise in the two-body channel only. They
take the form
\be
\Delta^{\mbox{\tiny 2-b}}_{\scs m=m_c} \G_{47}^{(1){\rm bare}} =
\fm{16}{81\ep} -\fm{4}{243} +\fm{264\pi^2-2186}{729}\ep + 2{\rm Re}\left[ b(z) + \ep \tilde b(z) \right] + {\mathcal O}(\ep^2).
\ee
Small-$z$ expansion of the function $b(z)$ has been given in Eq.~(3.9) of Ref.~\cite{Buras:2002tp},
while the large-$z$ expansion of ${\rm Re}\,b(z)$ can be found Eq.~(5.2) of Ref.~\cite{Misiak:2006ab}.
As far as $\tilde b(z)$ is concerned, we obtain the following expansions:
\bea
{\rm Re}\,\tilde b(z) &=&
\fm{1144}{729} - \fm{46}{243} \pi^2  - \fm{8}{243} L - \fm{2}{81} L^2
+ \fm{1}{z} \left(\fm{10957}{60750} + \fm{212}{2025} L + \fm{1}{15} L^2\right)\nnb\\[2mm]
&+& \fm{1}{z^2} \left( \fm{491839}{41674500} + \fm{134}{33075} L + \fm{2}{63} L^2\right)
+ {\mathcal O}\left( \fm{1}{z^3} \right),\nnb\\[3mm]
{\rm Re}\,\tilde b(z) &=&
\left( \fm{44}{3} - \fm{16}{9} \pi^2 - \fm{40}{9} \zeta_3 + \fm{16}{9} L - \fm{8}{9} L^2 \right) z
+ \left( \fm{304}{81} - \fm{128}{27} \ln 2 - \fm{32}{27} L\right) \pi^2 z^\f32\nnb\\[2mm] 
&+& \left( \fm{53}{3} - \fm{20}{27} \pi^2 + \fm{14}{3} L - \fm{32}{27} \pi^2 L + \fm{10}{9} L^2 - \fm{4}{9} L^3\right) z^2
- \fm{80}{27} \pi^2 z^\f52\nnb\\[2mm]
&+& \left( \fm{6830}{729} - \fm{292}{243} \pi^2 + \fm{80}{27} \zeta_3 + \fm{68}{243} L + \fm{64}{81} \pi^2 L
           - \fm{124}{27} L^2 + \fm{16}{9} L^3\right) z^3
+ \fm{88}{135} \pi^2 z^\f72\nnb\\[2mm]
&+& \left( \fm{1944727}{121500} - \fm{304}{405} \pi^2 + \fm{32}{9} \zeta_3 - \fm{17239}{2025} L - \fm{80}{27} L^2 + \fm{16}{9} L^3\right) z^4
+ \fm{272}{2835} \pi^2 z^\f92\nnb\\[2mm]
&+& \left( \fm{34017647}{833490} - \fm{1018}{189} \pi^2 + \fm{80}{9} \zeta_3 - \fm{113308}{3969} L
        - \fm{182}{27} L^2 + \fm{40}{9} L^3 \right) z^5 + {\mathcal O}\left(z^\f{11}{2}\right).
\eea
No explicit expressions for the expansions of $\tilde b(z)$ have so
far been published, even though this function must have been used for
UV renormalization in Ref.~\cite{Boughezal:2007ny}.

\end{document}